**Title:** Designing Optimal Distorted-Octahedra Superlattices for Strong Topological Hall Effect

Yiyan Fan,† Qinghua Zhang†, Jingdi Lu, Chuanrui Huo, Tianyang Wang, Qiao Jin, Ting Cui, Qianying Wang, Dongke Rong, Shiqing Deng, Lingfei Wang, Kuijuan Jin*, Jun Chen*, and Er-Jia Guo*

Y. Y. Fan, C. R. Huo, S. Q. Deng and Prof. J. Chen
Beijing Advanced Innovation Center for Materials Genome Engineering, Department of Physical Chemistry, University of Science and Technology Beijing, Beijing, 100083, China
E-mail: junchen@ustb.edu.cn

Q. H. Zhang, Q. Jin, T. Cui, Q. Y. Wang, D. K. Wang, Prof. K. J. Jin, and Prof. E. J Guo
Beijing National Laboratory for Condensed Matter Physics and Institute of Physics, Chinese Academy of Sciences, Beijing 100190, China
E-mail: kjjin@iphy.ac.cn and ejguo@iphy.ac.cn

J. D. Lu, and Prof. L. F. Wang
Hefei National Laboratory for Physical Sciences at Microscale, University of Science and Technology of China, Hefei 230026, China

T. Y. Wang
National Synchrotron Radiation Laboratory, University of Science and Technology of China

† These authors contribute equally to the manuscript.

**Abstract:**

Topologically protected spin states hold great promise for applications in next generation of memory circuits and spintronic devices. These intriguing textures typically emerge in bulk materials or heterostructures with broken inversion symmetry, accompanied by an enhanced Dzyaloshinskii-Moriya interaction (DMI). In this study, we successfully induced the topological Hall effect (THE) in atomically designed $(DyScO_3)_n/(SrRuO_3)_n$ ($D_nS_n$) superlattices over a significant range of temperatures (10~120K) and thicknesses (16~40nm). Using magnetic force microscopy (MFM), we observed the formation and stability of magnetic domains, such as topological skyrmions. By precisely controlling the interlayer thickness (*n*) and biaxial strain, we elucidated the mechanisms underlying the modulation and induction of magnetic topological states. Supporting evidence was provided by scanning transmission electron microscopy (STEM) and X-ray absorption spectroscopy (XAS), thereby lending further credence to our conclusions. These heterostructures offer a universal method for exploring topological phenomena driven by distorted octahedra, while enhancing the integrability and addressability of topologically protected functional devices.

**Main text**

1. **Introduction**

Topological spin textures in magnetic oxides are considered highly promising candidates for the development of future high-density, low-power, high-speed magnetic memories and logic circuits due to their high stability, minimal driving current and small dimensions.[1, 2] Theoretically, non-collinear spin textures, such as magnetic skyrmions, can be stabilized through dipole-dipole interactions,[3] Dzyaloshinskii-Moriya interactions (DMI),[4, 5] and frustrated exchange interactions.[6, 7] Among these mechanisms, DMI, which originates from asymmetric exchange interaction, is particularly effective in inducing orthogonal adjacent spin configurations in materials with strong spin–orbit coupling (SOC) and broken inversion symmetry.[8] The topological Hall effect (THE), which is closely related to these non-collinear spin configurations, generates a Berry phase in real space, thereby contributing an additional term to the Hall resistance.[9, 10] Given the significant challenges in directly observing nanoscale topological domain structures in real space, the THE has emerged as a classic hallmark indicative of the presence of topological spin textures.[11-14]

In 1994, Bogdanov and Hubert predicted the existence of topological spin textures in non-centrosymmetric magnetic crystals with bulk DMI or in asymmetric magnetic multilayers with interfacial DMI.[15] $SrRuO_3$ (SRO), a $3d$ transition metal oxide with strong spin-orbit coupling, serves as an ideal platform for studying magnetic topological structures. Previous studies by Lu and Wang *et al*. demonstrated the induction of the THE in single-layer SRO through size effects and the introduction of oxygen defects, leading to asymmetric center structures within the bulk material.[16, 17] Matsuno *et al*. induced a pronounced THE in $SrRuO_3$/$SrIrO_3$ heterostructures by incorporating $5d$ transition metal oxides with stronger SOC.[18] Additionally, researchers have constructed heterostructures of SRO with functional oxides such as ferroelectrics ($BaTiO_3$/$PbTiO_3$), ferromagnetics ($LaSrMnO_3$), and multiferroics ($BiFeO_3$), all of which exhibited emergent THE.[17, 19-23] These studies attribute the emergent THE to enhanced interfacial DMI, which originates from atomic displacement, the interplay of magnetic crystal anisotropy, and exchange coupling. Besides above methods, octahedral tilting, a traditional lattice degree of freedom, can precisely tailor the lattice structure and symmetry in SRO.[24] Specifically, the geometry of the Ru-O bonds significantly influences SRO's performances, including metal-insulator transitions, superconductivity, electrocatalysis and novel magnetoelectric transport.[25-27] A thorough exploration of the universal principles and the influence of octahedral tilt and distortion on triggering the THE in SRO is essential and pressing, given that this aspect remains unexamined.

Theoretical studies have shown that the orthorhombic (with octahedral tilting) and tetragonal (without octahedral tilting) structures of SRO are nearly degenerate in energy. This energetic proximity allows for facile manipulation of the octahedral structure within a unit cell via external fields such as temperature, strain and magnetic fields.[28] As shown in Figure 1(a), in addition to the mismatch strain between the film and substrate, the symmetrical structure of the substrate can impose additional geometric constraints. Unlike the cubic symmetry of the $SrTiO_3$ substrate, which suppresses octahedral rotation, the oxygen octahedra in SRO can replicate the tilting characteristics of the $DyScO_3$ (DSO) substrate, one of the orthorhombic scandate materials with the largest octahedral tilting angle.[29, 30] In this work, we selected the DSO interlayer to induce octahedral distortion in SRO and constructed superlattices through repeated stacking to enhance interfacial DMI. Consequently, additional THE and discontinuous magnetic topological domains were observed in Hall and magnetic force microscopy (MFM) measurements. High-resolution scanning transmission electron microscopy (STEM) and X-ray absorption spectroscopy (XAS) further confirmed that these phenomena primarily originated from the octahedral distortion propagated through the heterointerface. Lastly, the THE completely disappears when the biaxial tensile strain is applied, even though the DSO interlayer still exists. This work provides a specific and reproducible guideline for inducing distorted octahedra and THE in multilayer heterostructures from the perspective of intercalation material and substrate template selection.

## 2. Results and Discussion

Figure 1(b) illustrates the principal designed concept, in which the continuous periodicity at the atomic scale in the superlattice film disrupts the lattice symmetry of bulk SRO and enhances the interfacial DMI. To explore the optimal effective depth of octahedral tuning near the interface, we grew a series of $(DSO)_n/(SRO)_n$ ($D_nS_n$) superlattice (SLs) samples with varying interlayer thicknesses, where $n$ represents the number of unit cells. We keep the thicknesses of DSO and SRO are the same. Figure 1(c) presents the X-ray diffraction θ-2θ scans of $D_nS_n$ SLs grown on STO substrates, featuring the high-quality epitaxial single-crystal peaks and coherent superlattice diffraction peaks. As the interlayer thickness increases, the positive-order diffraction peaks continuously shift to higher angles. This observation is consistent with the formula for the thickness as a function of the diffraction peak position: $d=(m_1-m_2)\lambda/2(\sin\theta_{m1}-\sin\theta_{m2})$, where $m$ is the order of the superlattice peak, $\lambda$ is the wavelength of the X-ray beam, θ is position of diffraction peak, and $d$ is the repeated thickness. The reciprocal space mapping (RSM) in the inset indicates that the $D_5S_5$ samples are fully strained on the substrate due to the repeated clamping from the two materials. RSMs on other samples were shown in Figure S2, Supporting Information. And then we measured the X-ray reflectivity (XRR), where the width of the oscillation peaks is inversely proportional to the thickness of the sample. For instance, following a specific fitting procedure, the $D_5S_5$ superlattice exhibits periodic density with a repeating unit of approximately 4 nm and a periodicity of 10 times (Figure S1, Supporting Information). Moreover, Figure 1(d) shows cross-sectional high-angle annular dark field scanning transmission electron microscopy (HDAAF-STEM) of the representative $D_5S_5$ along the $[010]_{pc}$ zone axis. Since the atomic intensity is approximately proportional to $Z^{1.7}$ (where $Z$ is the atomic number), this image reveals the periodic structure of alternating SRO and DSO at the atomic scale. For better contrast, we presented the average intensity mapping of the A-site atoms (Ru and Dy) from Figure 1(d) in Figure 1(e). All the evidence demonstrates that we have obtained high-quality, continuously strained superlattices as expected.

To investigate the magnetoelectric characteristics of the superlattices regulated by structure, we measured the temperature-dependent resistance using DC transport measurements, as illustrated in Figure 2(a). Bulk DSO exhibits excessively high resistance, making it unmeasurable resistance (over 1 MΩ) at room temperature, whereas bulk SRO displays metallic behavior. For $n > 4$, the metallic characteristics are predominantly attributed to the SRO layers, due to the minimal charge transfer in this regime. In contrast, for $n < 4$, the samples exhibit semi-metallic behavior with a metal-insulator transition temperature ranging from 20 to 40 K. Considering that single SRO and DSO layers are insulating at these equivalent ultrathin thicknesses, the enhanced metallicity is likely due to the removal of electrical dead layer in SLs.[31] Additionally, the resistance increase below the transition temperature can be attributed to the size effect or interface localization, while the shift in transition temperature with decreasing thickness is determined by the competition between localization and delocalization mechanisms. [32, 33] Notably, all curves display an inflection point strongly correlated with the ferromagnetic Curie temperature.[34, 35] The derivative curves of resistance with respect to temperature indicate that the transition temperature decreases from 130 K to 100 K as the superlattice changes from $D_7S_7$ to $D_2S_2$ (Figure S3, Supporting Information). These findings suggest that the magnetic transport properties of the superlattice samples are predominantly influenced by SRO and precisely controlled by the interlayer thickness.

As anticipated, the THE exhibits a pronounced dependence on the repeated unit of the superlattices. Theoretically, the total Hall effect can be decomposed as $\rho_{xy} = \rho_{OHE} + \rho_{AHE} + \rho_{THE}$, where $\rho_{OHE}$, $\rho_{AHE}$ and $\rho_{THE}$ correspond to the ordinary Hall resistivity, anomalous Hall resistivity, and topological Hall resistivity, respectively. These components can be expressed as $\rho_{OHE} = R_0 H$, $\rho_{AHE} = R_s M$, where $R_0$、$R_s$、$H$ and $M$ represent the ordinary Hall coefficient, anomalous Hall coefficient, out-of-plane magnetic field, and magnetization, respectively.[36-38] As shown in Figure 2(b), the Hall resistivity, ($\rho_{xy}$-R$_0$H)-H, for samples with varying thickness at different temperatures have been corrected by subtracting the contribution from the ordinary Hall effect through fitting the linear slope. For samples with $n < 5$, the ($\rho_{xy}$-R$_0$H)-H curves exhibit a bimodal feature near the coercive field, indicating the presence of THE. Conversely, for samples with $n > 5$, the bimodal feature vanishes, resembling the contribution from the anomalous Hall effect.

Subsequently, we utilized the out-of-plane M–H curves to fit the contributions of the anomalous Hall effect (AHE) and precisely estimate the THE.[18, 39] As depicted by the green line in Figure 2(c), the pure topological Hall resistivity as a function of magnetic field ($\rho_{THE}$–H) was obtained. When the magnetic field increased from −5 T to −2 T, the curve was dominated by the ferromagnetic state with negative magnetization, characterized by finite anomalous Hall conductivity and negligible topological Hall conductivity. As the magnetic field further increased from −2 T to 5 T, the topological Hall resistivity abruptly increased at 1.6 T and gradually decreased to zero at 3.5 T, coinciding with the saturation field of the M–H curve. These observations suggest that specific chiral spin structures are induced as the ferromagnetic spins begin to reverse under the influence of the magnetic field. The coexistence of magnetic hysteresis and THE suggests indicates the presence of both the ferromagnetic phase and non-collinear spin configurations.

Next, we applied the same procedure to quantify the THE in samples with n ≤ 5 at $T = 10$ K by calculating the density of magnetic topological domains. Under strong exchange coupling between charge carriers and topological spin textures, $\rho_{THE}$ can be formulated as $\rho_{THE}=P_s R_0 n_{sk} h/e$, where $P_s$、$n_{sk}$、h and e denote the spin polarization of the charge carriers, the density of magnetic topological domains, the quantum of flux, and the elementary charge, respectively.[40-42] For SRO, $P_s$ is typically −9.5±5%.[18, 43] Based on this, we summarized the evolution of $R_0$ and $n_{sk}$ with interlayer thickness. The rapid decay of $n_{sk}$ with increasing $n$ (Figure 2d) mirrors the trend of the topological Hall resistivity. The reasons for these phenomena will be detailed explained later. Figure 2(e) presents a summary phase diagram of the topological Hall resistivity as a function of interlayer thickness and temperature. It is evident that a strong THE exists in the region where $n$ is 2~5 unit cells and can be maintained over a broad temperature range of $T$ =10~100 K. This temperature range is broader than the reports of the existing SRO.[17, 18]

Magnetic force microscopy (MFM) is a widely used technique for observing various magnetic structures in real space.[44, 45] Given that the out-of-plane magnetization and the magnitude of the topological Hall resistivity are within detective limits of the equipment, we obtained MFM images of the $D_2S_2$ sample over a magnetic field range of 0–5 T (Figure S4, Supporting Information). However, the poor contrast in the MFM images, due to the mixture of morphology, stray fields, and domain signals, makes it challenging to observe magnetic topological signals in the static mode at ultra-low temperatures.[46] To address this issue, we performed pixel-by-pixel subtraction on the images within the magnetic field range near the Hall peaks (1.5–1.7 T), following previously established methods.[17, 45] The processed MFM image, which shows clear contrast between the magnetic domain structure (yellow) and the background (blue). Based on the equivalent diameter of the various configurations, we classified the magnetic domains in the MFM images into three categories: Type 1, Type 2, and Type 3. Type 1 consists of uniformly sized, quasi-circular domains with dimensions ranging from 20 to 80 nm,[17] resembling magnetic skyrmions and accounting for 45.5% of all observed textures, (see the statistical histogram in Figure S4(f), supporting information). Type 2 includes irregular skyrmion clusters or bubble domains with dimensions between 80 and 200 nm,[47] representing approximately 36.4% of the structures. Furthermore, Type 3 comprises larger, chain-like domains with dimensions from 200 to 400 nm, which may correspond to bubble domains formed by adjacent skyrmions or traditional continuous domains driven by the magnetic field, accounting for 18.1% of the structures. This analysis confirms the presence and variety of topological textures in the $D_2S_2$ sample, thereby substantiating the genuine origin of the THE, as opposed to the superposition of multi-channel anomalous Hall effects.[48]

Using HADDF and ABF-STEM with atomic resolution, we uncovered the microscopic mechanisms underlying the induction of stable THE within the multilayer heterostructures. Consistent with the initial design concept, exploring the propagation of tilting and distorted octahedra in the repeated stacking heterojunctions is crucial. In Figure 3(a), along the $[110]_{pc}$ zone axis, we obtained the HADDF-STEM images of the $D_2S_2$ sample with the highest $n_{sk}$. The color contrast in the images clearly distinguishes the duplicate layers composed of DSO and SRO, each with a repeated thickness of 2-unit cells. Subsequently, we further acquired ABF-STEM images of the same region to determine the precise positions of the oxygen atoms within the oxygen octahedra. Compared to the $[100]_{pc}$ zone axis, STEM images along the $[110]_{pc}$ zone axis facilitate intuitive quantification of the tilt and distortion of the $BO_6$ octahedra (B = Ru, Sc). Ideally, when a 2-unit cell layer of SRO is sandwiched between the upper and lower layers of DSO, the $RuO_6$ octahedron should align with the characteristics of $ScO_6$. In our work, we calculated the actual rigid tilting angles of the oxygen octahedra based on the geometric relationships between atoms, as illustrated in the schematic in Figure 3(c). The exact positions of Sc, Ru, and O atoms, determined by Gaussian fitting of atomic columns, are labeled in the ABF-STEM images. Figure 3(c) shows the variation of the octahedral tilting angle with atomic layer position, with the dashed line denotes the theoretical value for the bulk $RuO_6$. Apparently, the $RuO_6$ octahedra in the $D_2S_2$ samples exhibit a larger tilting angle (8.2°) than that in the tetragonal bulk, which is 1.34 times the theoretical value.[25] In addition to the rigid tilting described above, we observed inner distortions accompanied by vertical shifts of the central B-position atoms. Based on precise atomic positions and geometric relationships, the bond angles between O-B-O, which depend on the atomic layer positions, are summarized in Figure 3(d). The maximum distorted angle is 168°, significantly less than the theoretical value of 180°. These results indicate that both rigid tilting and inner distortions are present in the designed samples, which are the primary reasons for the introduction of broken symmetry, enhanced DMI, and the strong THE.

The transport results in Figure 2 show that the THE is highly dependent on the interlayer thickness of the superlattice. Specifically, the THE disappears when $n > 7$. We attribute this to two main factors. First, studies on related heterostructures have demonstrated that structural symmetry propagation is primarily confined within 2~4 unit cells near the interface.[49-52] When n ⩽ 5, all SRO unit cells are influenced by the DSO material from the adjacent layer, leading to enhanced tilting and distortion of the $RuO_6$ octahedra, which in turn increases the DMI and induces the THE. However, when n > 7, only the unit cells near the interface inherit the structural symmetry from DSO, while the middle layers retain bulk characteristics. The superposition of these two effects is not conducive to induce THE, which also explains why the intensity of the topological Hall peaks decreases with increasing *n* in the range of *n* < 5. Second, as *n* increases, the total thickness of SLs rises from 16 to 80 nm, leading to strain relaxation exerted from the substrates. In this scenario, the SRO layers near the surface primarily withstands the tensile strain from the neighboring DSO layer, which has been reported to suppress the tilting and distortion of the octahedra.[24] Figure 3(d) illustrates the variation of the out-of-plane lattice parameter as a function of interlayer thickness. As *n* increases, the macroscopic lattice parameters decrease and gradually approach the *c*-value of bulk SRO, indicating that the unit cells far from and close to the substrate experience different in-plane strains. The broadening of the film peak in the $D_7S_7$ sample along the in-plane direction in the RSM further corroborates this point. Finally, considering the variation of octahedral tilting and strain with thickness, we constructed an intuitive model diagram to better illustrate above viewpoints (Figure S5, Supporting Information).

To avoid the interference of contradictory strain and relaxation phenomena between the substrate and interlayer, we further deposited the $D_2S_2$ superlattice on $KTaO_3$ (KTO) substrates, which have the largest in-plane parameters compared to bulk SRO and DSO (see Figure 4(a)). The structural characterizations of the $D_2S_2$ superlattices on KTO and STO substrates (Figure S6, supporting Information), confirming that the high-quality epitaxial films are fully subjected to biaxial tensile and compressive strains, respectively. Similar to samples on STO substrates with $n>7$, the $D_2S_2$ samples under tensile strain exhibit only typical anomalous Hall features, without any contribution from the THE. This is in sharp contrast to samples of the same thickness under compressive strain. These results indicate that the THE, driven by oxygen octahedral distortions, can only be induced by precisely co-tuning the interplay thickness and biaxial strain, consistent with our hypothesis.

Combining synchrotron X-ray absorption spectra, we further elucidated the electronic structural origin of the THE. The O $K$-edge XAS spectra of 4d transition-metal oxides represent transitions from the O 1$s$ states to unoccupied transition-metal 4$d$ and 5$s$/5$p$ states, as well as to other conduction-band states via the hybridization with the O 2$p$ states.[53, 54] Figure 4(c) shows the XAS spectra of the two samples measured at an incident X-ray angle of 40° relative to the sample surface. The peaks near 529 eV are attributed to the absorption of the coherent and incoherent parts of the Ru 4$d$ $t_{2g}$ states, while transitions to the Ru 4$d$ $e_g$ states occur within the energy range of 533 eV.[55, 56] These peak positions for the Ru-4$d$ state basically coincide with those in bulk SRO, with differences not exceeding 0.5 eV, indicating that Ru ions in the superlattices mainly exist in the +4-valence state with negligible charge transfer at interfaces. Given the rich information contained in the hybrid peaks between Ru 4$d$ and O 2$p$, such as ferromagnetic origin and structural distortion, we conducted detailed analysis of the charge distribution and transfer within the Ru 4d orbitals. Compared to the sample under $\varepsilon = +1.5\%$, when $\varepsilon = -0.7\%$, the intensity of the hybrid peak between O 2$p$ and Ru 4$d$ $t_{2g}$ increases, while that between O 2$p$ and Ru 4$d$ $e_g$ decreases. This corresponds to an increase in the number of empty states in the Ru 4$d$ $t_{2g}$ orbitals and a decrease in the number of empty states in the Ru 4$d$ $e_g$ orbitals. Concurrently, the reduction of the crystal field splitting energy ($\Delta C_F$) facilitates electron migration from the Ru 4$d$ $t_{2g}$ orbital to the $e_g$ orbital, resulting in a high spin state in the SRO. The measured magnetization of 4 $\mu_B$/Ru on the STO substrate also corroborates the electron states of the Ru 4d orbitals discussed above (Figure S7, Supporting Information). In contrast, the saturation magnetization of approximately 2 μB/Ru in the tensile strain sample is consistent with the low or medium spin states produced by weak octahedral distortion in tetragonal or orthorhombic structures.[26] Figure S8 provides a detailed schematic of the electronic structure in the Ru 4$d$ orbitals for both distorted and bulk octahedra.

Figure 4(d) shows the X-ray linear dichroism spectra calculated from $I_{ab}$ - $I_c$, where $I_{ab}$ and $I_c$ directly reflect the unoccupied electron states in the in-plane and out-of-plane orbitals, respectively. The difference between them directly indicates the dominance of electrons in either the in-plane or out-of-plane orbital components.[57] According to the electronic structure of the Ru 4d orbital, electrons primarily occupy the lower energy orbitals of $d_{xz}$ and $d_{yz}$ at ε = +1.5%, while the occupancy of the $d_{xy}$ orbital increases with electron transfer when the ε = -0.7%. These scenarios are reflected in the XAS spectra, near 529 eV, the intensity of $I_{ab}$-$I_c$ is significantly higher in the tensile strain sample than that in the compressive strain sample.[58] Additionally, near 533 eV, a larger $I_{ab}$-$I_c$ intensity corresponds to an increase in the occupancy of the $d_{3z^2-r^2}$ orbitals under compressive strain. These results demonstrate that the biaxial compressive strain facilitates the propagation of distorted octahedra in SRO without thickness relaxation, which is the origin of high spin states and strong THE.

## 3. Conclusion

In summary, we observed a pronounced THE by designing superlattices of repeated DSO/SRO at the atomic scale. Within these multilayer heterostructures, we explored the modulated rules of octahedral behavior on the electrical transport performance and magnetic spin states in transition metal oxides. By selectively manipulating two variables, interlayer thickness and the biaxial strain, we have established a universal design principle: high-density, stable, and non-volatile magnetic topological structures can be realized when multilayer heterostructures are subjected to fully compressive strain and the interlayer thickness is controlled to 2~5-unit cells. This result, leveraging the propagation of structural symmetry, provides a precedent for precisely controlling magnetic topological configurations driven by distorted octahedra in non-chiral magnets. This is crucial for the multidimensional research into future topological electronics. Moreover, this all-oxide skyrmion system can be flexibly and epitaxially grown with other functional oxide heterostructures, offering a fertile playground for exploring emergent properties applicable to the fields of memory devices, sensors, and spintronic devices.

## 4. Experimental Section

**PLD synthesis of superlatttices:** We fabricated $D_nS_n$ superlattices with various thicknesses on (100)-oriented STO substrates using pulsed laser deposition. Sintered SRO and DSO ceramics served as ablation targets to produce thin films with precise chemical stoichiometry. To facilitate the alternating growth of the two materials during superlattice fabrication, we optimized the growth conditions to be identical parameters for both materials while maintaining their individual high quality and single-crystallinity. During fabrication, the laser fluence was set at 1.5-2 J/cm$^2$, the repetition rate was 2 Hz, and the oxygen pressure was maintained at 0.1 Torr. All films were grown at 750 °C and subsequently cooled to room temperature at a rate of −10 °C/min after laser irradiation. The thickness of individual layers was precisely controlled by counting the number of laser pulses, with average growth rates of 110 pulses/u.c. for DSO and 50 pulses/u.c. for SRO.

**Structural characterizations and physical property measurement:** X-ray diffraction θ–2θ scans and reflectivity measurements for $D_nS_n$ superlattices were performed using a PANalytical X'Pert3 MRD diffractometer. Reciprocal space mappings were conducted at the 1W1A beamline of Beijing Synchrotron Radiation Facility(BSRF). Longitudinal ($\rho_{xx}$) and Hall ($\rho_{xy}$) resistivity hysteresis loops were measured using a Quantum Design 9-T Physical properties measurement system(PPMS) with magnetic fields were applied normal to the sample surfaces. The measurements were conducted using the standard van der Pauw geometry with wire-bonding techniques. *M-H* hysteresis loops up to 5 T were measured using a Quantum Design superconducting quantum interference device (SQUID) magnetometer at $T = 10$ K with fields applied perpendicular to the sample surface.

**STEM measurement and analysis:** Cross-sectional STEM images of DnSn superlattices were obtained along the pseudo-cubic [100]/[110] zone axis using an aberration-corrected FEI Titan Themis G2 microscope. While HADDF-STEM excel in visualizing the periodicity of heterogeneous structures, ABF-STEM facilitate the direct observation of oxygen atoms, which is crucial for analyzing oxygen octahedral tilt. The precise positions of Ru, Sc and O ions were determined by fitting the intensity peaks with a Gaussian function. The tilting and distortion angles were calculated using the bond lengths (Ru-O/Sc-O) and trigonometric function in the ABF-STEM. Each data point represents the average tilting and distortion angles from over 20-unit cells at the same layer, with error bars derived from the standard deviation (SD) value.

**Magnetic force microscopy:** MFM measurements performed using a commercially available scanning probe microscope (Attocube system) equipped with a 9-T superconducting magnet and a Helium refrigeration system. Commercial magnetic probes (Nanosensor, PPP-MFMR) were used for all MFM measurements, scanning at a constant height. The magnetic-field dependent MFM measurements were conducted with the external magnetic field applied perpendicularly to the film surface.

**XAS and XLD measurements:** Elemental-specific XAS measurements were performed in the total electron yield (TEY) mode for the O $K$-edges at the 4B9B beamline of the Beijing Synchrotron Radiation Facility (BSRF). The sample's scattering plane was oriented at 40° and 90° relative to the incident X-ray beam, with the difference was calculated by $I_{ip} - I_{oop}$. When the x-ray beam is perpendicular to the surface plane, the XAS signal reflects the occupied states in the in-plane orbitals. At a 40° angle between the X-ray beam and the surface plane, the XAS signal captures information from both in-plane and out-of-plane orbitals. For simplified analysis, the unoccupied in-plane orbital states are proportional to $I_{ip} = I_{90°}$, while the unoccupied out-of-plane orbital states are derived from $I_{oop} = (I_{90°} - I_{40°} \cdot \sin^2 40°)/\cos^2 30°$.

**Supporting information**

Supporting Information is available from the Wiley Online Library or from the author.

**Acknowledgements**

This work was supported by the Beijing Natural Science Foundation (Grant No. JQ24002), the National Key Basic Research Program of China (Grant Nos. 2020YFA0309100), the National Natural Science Foundation of China (Grant No. 22235002, U22A20263, 52250308), the CAS Project for Young Scientists in Basic Research (Grant No. YSBR-084).

**Conflict of Interest**

The authors declare no conflict of interest.

**Keywords:** distorted oxygen octhedral, interfacial Dzyaloshinskii–Moriya interaction, multilayer heterostructures, topological Hall effect, magnetic topological textures

Received: ((will be filled in by the editorial staff))

Revised: ((will be filled in by the editorial staff))

Published online: ((will be filled in by the editorial staff))

**References**

[1] A. Fert, N. Reyren, V. Cros, *Nat. Rev. Mater.* **2017**, 2, 17031.

[2] G. Yu, P. Upadhyaya, Q. Shao, H. Wu, G. Yin, X. Li, C. He, W. Jiang, X. Han, P. K. Amiri, K. L. Wang, *Nano Lett.* **2016**, 17, 261-268.

[3] M. Ezawa, *Phys. Rev. Lett.* **2010**, 105, 197202.

[4] T. Moriya, *Phys. Rev.* **1960**, 120, 91-98.

[5] U. K. R¨oßler, A. N. Bogdanov, *Nature* **2006**, 442, 797-801.

[6] A. O. Leonov, M. Mostovoy, *Nat. Commun.* **2015**, 6, 8275.

[7] X. Zhang, J. Xia, Y. Zhou, X. Liu, H. Zhang, M. Ezawa, *Nat. Commun.* **2017**, 8, 1717.

[8] S. Rohart, A. Thiaville, *Phys. Rev. B* **2013**, 88, 184422.

[9] D. Maccariello, W. Legrand, N. Reyren, K. Garcia, K. Bouzehouane, S. Collin, V. Cros, A. Fert, *Nat. Nanotechnol.* **2018**, 13, 233-237.

[10] N. Nagaosa, Y. Tokura, *Nat. Nanotechnol.* **2013**, 8, 899-911.

[11] X. Niu, B.-B. Chen, N. Zhong, P.-H. Xiang, C.-G. Duan, *J. Phys.: Condens.Matter* **2022**, 34, 244001.

[12] Y. Li, N. Kanazawa, X. Z. Yu, A. Tsukazaki, M. Kawasaki, M. Ichikawa, X. F. Jin, F. Kagawa, Y. Tokura, *Phys.l Rev. Lett.* **2013**, 110, 117202.

[13] C. Sürgers, G. Fischer, P. Winkel, H. v. Löhneysen, *Nat. Commun.* **2014**, 5, 3400.

[14] S. X. Huang, C. L. Chien, *Phys. Rev. Lett.* **2012**, 108, 267201.

[15] A. Bogdanov, A. Hubert, *J. Magn. Magn. Mater.* **1994**, 138, 255-269.

[16] J. Lu, L. Si, Q. Zhang, C. Tian, X. Liu, C. Song, S. Dong, J. Wang, S. Cheng, L. Qu, K. Zhang, Y. Shi, H. Huang, T. Zhu, W. Mi, Z. Zhong, L. Gu, K. Held, L. Wang, J. Zhang, *Adv. Mater.* **2021**, 33, 2102525.

[17] L. Wang, Q. Feng, Y. Kim, R. Kim, K. H. Lee, S. D. Pollard, Y. J. Shin, H. Zhou, W. Peng, D. Lee, W. Meng, H. Yang, J. H. Han, M. Kim, Q. Lu, T. W. Noh, *Nat. Mater.* **2018**, 17, 1087-1094.

[18] J. Matsuno, N. Ogawa, K. Yasuda, F. Kagawa, W. Koshibae, N. Nagaosa, Y. Tokura, M. Kawasaki, *Sci. Adv.* **2016**, 2, e1600304.

[19] X. Yao, C. Wang, E.-J. Guo, X. Wang, X. Li, L. Liao, Y. Zhou, S. Lin, Q. Jin, C. Ge, M. He, X. Bai, P. Gao, G. Yang, K.-j. Jin, *ACS Appl. Mater. Interfaces* **2022**, 14, 6194-6202.

[20] H. Wang, Y. Dai, Z. Liu, Q. Xie, C. Liu, W. Lin, L. Liu, P. Yang, J. Wang, T. V. Venkatesan, G. M. Chow, H. Tian, Z. Zhang, J. Chen, *Adv. Mater.* **2020**, 32, 1904415.

[21] E. Skoropata, J. Nichols, J. M. Ok, R. V. Chopdekar, E. S. Choi, A. Rastogi, C. Sohn, X. Gao, S. Yoon, T. Farmer, R. D. Desautels, Y. Choi, D. Haskel, J. W. Freeland, S. Okamoto, M. Brahlek, H. N. Lee, *Sci. Avd.* **2020**, 6, eaaz3902.

[22] I. Lindfors-Vrejoiu, M. Ziese, *Phys. Status Solidi B* **2016**, 254, 1600556.

[23] S. D. Seddon, D. E. Dogaru, S. J. R. Holt, D. Rusu, J. J. P. Peters, A. M. Sanchez, M. Alexe, *Nat. Commun.* **2021**, 12, 2007.

[24] W. Lu, W. Song, P. Yang, J. Ding, G. M. Chow, J. Chen, *Sci. Rep.* **2015**, 5, 10245.

[25] Z. Liao, M. Huijben, Z. Zhong, N. Gauquelin, S. Macke, R. J. Green, S. Van Aert, J. Verbeeck, G. Van Tendeloo, K. Held, G. A. Sawatzky, G. Koster, G. Rijnders, *Nat. Mater.* **2016**, 15, 425-431.

[26] S. G. Jeong, G. Han, S. Song, T. Min, A. Y. Mohamed, S. Park, J. Lee, H. Y. Jeong, Y. M. Kim, D. Y. Cho, W. S. Choi, *Adv. Sci.* **2020**, 7, 2001643.

[27] S. G. Jeong, S. W. Cho, S. Song, J. Y. Oh, D. G. Jeong, G. Han, H. Y. Jeong, A. Y. Mohamed, W.-s. Noh, S. Park, J. S. Lee, S. Lee, Y.-M. Kim, D.-Y. Cho, W. S. Choi, *Nano Lett.* **2024**, 24, 7979-7986.

[28] A. Herklotz, K. Dörr, *The Eur. Phys. J. B* **2015**, 88, 60.

[29] R. P. Liferovich, R. H. Mitchell, *J. Solid State Chem.* **2004**, 177, 2188-2197.

[30] R. Uecker, B. Velickov, D. Klimm, R. Bertram, M. Bernhagen, M. Rabe, M. Albrecht, R. Fornari, D. G. Schlom, *J. Cryst. Growth* **2008**, 310, 2649-2658.

[31] J. Nichols, X. Gao, S. Lee, T. L. Meyer, J. W. Freeland, V. Lauter, D. Yi, J. Liu, D. Haskel, J. R. Petrie, E.-J. Guo, A. Herklotz, D. Lee, T. Z. Ward, G. Eres, M. R. Fitzsimmons, H. N. Lee, *Nat. Commun.* **2016**, 7, 12721.

[32] J. Shin, A. Y. Borisevich, V. Meunier, J. Zhou, E. W. Plummer, S. V. Kalinin, A. P. Baddorf, *ACS Nano* **2010**, 4, 4190-4196.

[33] G. Herranz, V. Laukhin, F. Sánchez, P. Levy, C. Ferrater, M. V. García-Cuenca, M. Varela, J. Fontcuberta, *Phys. Rev. B* **2008**, 77, 165114.

[34] M. E. Fisher, J. S. Langer, *Phys. Rev. Lett.* **1968**, 20, 665-668.

[35] X. Shen, X. Qiu, D. Su, S. Zhou, A. Li, D. Wu, *J. Appl. Phys.* **2015**, 117, 015307.

[36] R. Mathieu, A. Asamitsu, H. Yamada, K. S. Takahashi, M. Kawasaki, Z. Fang, N. Nagaosa, Y. Tokura, *Phys. Rev. Lett.* **2004**, 93, 016602.

[37] S. Shimizu, K. S. Takahashi, M. Kubota, M. Kawasaki, Y. Tokura, Y. Iwasa, *Appl. Phys. Lett.* **2014**, 105, 163509.

[38] G. Koster, L. Klein, W. Siemons, G. Rijnders, J. S. Dodge, C.-B. Eom, D. H. A.

Blank, M. R. Beasley, *Rev. Mod. Phys.* **2012**, 84, 253-298.

[39] A. Soumyanarayanan, M. Raju, A. L. Gonzalez Oyarce, A. K. C. Tan, M.-Y. Im, A. P. Petrović, P. Ho, K. H. Khoo, M. Tran, C. K. Gan, F. Ernult, C. Panagopoulos, *Nat. Mater.* **2017**, 16, 898-904.

[40] F. D. M. Haldane, *Phys. Rev. Lett.* **2004**, 93, 206602.

[41] D. Xiao, M.-C. Chang, Q. Niu, *Rev. Mod. Phys.* **2010**, 82, 1959-2007.

[42] S. Chakraverty, T. Matsuda, H. Wadati, J. Okamoto, Y. Yamasaki, H. Nakao, Y. Murakami, S. Ishiwata, M. Kawasaki, Y. Taguchi, Y. Tokura, H. Y. Hwang, *Phys. Rev. B* **2013**, 88, 220405.

[43] D. C. Worledge, T. H. Geballe, *Phys. Rev. Lett.* **2000**, 85, 5182.

[44] H. Zhou, Z. Wang, Y. Hou, Q. Lu, *Ultramicroscopy* **2014**, 147, 133-136.

[45] K.-Y. Meng, A. S. Ahmed, M. Baćani, A.-O. Mandru, X. Zhao, N. Bagués, B. D. Esser, J. Flores, D. W. McComb, H. J. Hug, F. Yang, *Nano Lett.* **2019**, 19, 3169-3175.

[46] A. Yagil, A. Almoalem, A. Soumyanarayanan, A. K. C. Tan, M. Raju, C. Panagopoulos, O. M. Auslaender, *Appl. Phys. Lett.* **2018**, 112, 192403.

[47] X. Yu, Y. Tokunaga, Y. Taguchi, Y. Tokura, *Adv. Mater.* **2016**, 29, 1603958.

[48] P. Roy, A. Carr, T. Zhou, B. Paudel, X. Wang, D. Chen, K. T. Kang, A. Pateras, Z. Corey, S. Lin, J. X. Zhu, M. V. Holt, J. Yoo, V. Zapf, H. Zeng, F. Ronning, Q. Jia, A. Chen, *Adv. Elect. Mater.* **2023**, 9, 2300020.

[49] A. Sahoo, P. Padhan, W. Prellier, *ACS Appl. Mater. Interfaces* **2017**, 9, 36423-36430.

[50] S. G. Jeong, T. Min, S. Woo, J. Kim, Y.-Q. Zhang, S. W. Cho, J. Son, Y.-M. Kim, J. H. Han, S. Park, H. Y. Jeong, H. Ohta, S. Lee, T. W. Noh, J. Lee, W. S. Choi, *Phys. Rev. Lett.* **2020**, 124, 026401.

[51] Y. Choi, Y. Z. Yoo, O. Chmaissem, A. Ullah, S. Kolesnik, C. W. Kimball, D. Haskel, J. S. Jiang, S. D. Bader, *Appl. Phys. Lett.* **2007**, 91, 022503.

[52] A. V. Boris, Y. Matiks, E. Benckiser, A. Frano, P. Popovich, V. Hinkov, Wochner, M. Castro-Colin, E. Detemple, V. K. Malik, C. Bernhard, T. Prokscha, A. Suter, Z.

Salman, E. Morenzoni, G. Cristiani, H.-U. Habermeier, B. Keimer, *Sci. Adv* **2011**, 332, 937-940.

[53] Y. K. Wakabayashi, M. Kobayashi, Y. Takeda, M. Kitamura, T. Takeda, R. Okano, Y. Krockenberger, Y. Taniyasu, H. Yamamoto, *Phys. Rev. Mater.* **2022**, 6, 094402.

[54] D. H. Kim, E. Lee, H. W. Kim, S. Kolesnik, B. Dabrowski, C.-J. Kang, M. Kim, B. I. Min, H.-K. Lee, J. Y. Kim, J. S. Kang, *Phys. Rev. B* **2015**, 91, 075113.

[55] M. Takizawa, D. Toyota, H. Wadati, A. Chikamatsu, H. Kumigashira, A. Fujimori, M. Oshima, Z. Fang, M. Lippmaa, M. Kawasaki, H. Koinuma, *Phys. Rev. B* **2005**, 72, 060404.

[56] K. Fujioka, J. Okamoto, T. Mizokawa, A. Fujimori, *Phys. Rev. B* **1997**, 56, 6380.

[57] S. Lin, Q. Zhang, X. Sang, J. Zhao, S. Cheng, A. Huon, Q. Jin, S. Chen, S. Chen, W. Cui, H. Guo, M. He, C. Ge, C. Wang, J. Wang, M. R. Fitzsimmons, L. Gu, T. Zhu, K. Jin, E.-J. Guo, *Nano Lett.* **2021**, 21, 3146-3154.

[58] F. Frati, M. O. J. Y. Hunault, F. M. F. de Groot, *Chem. Rev.* **2020**, 120, 4056-4110.

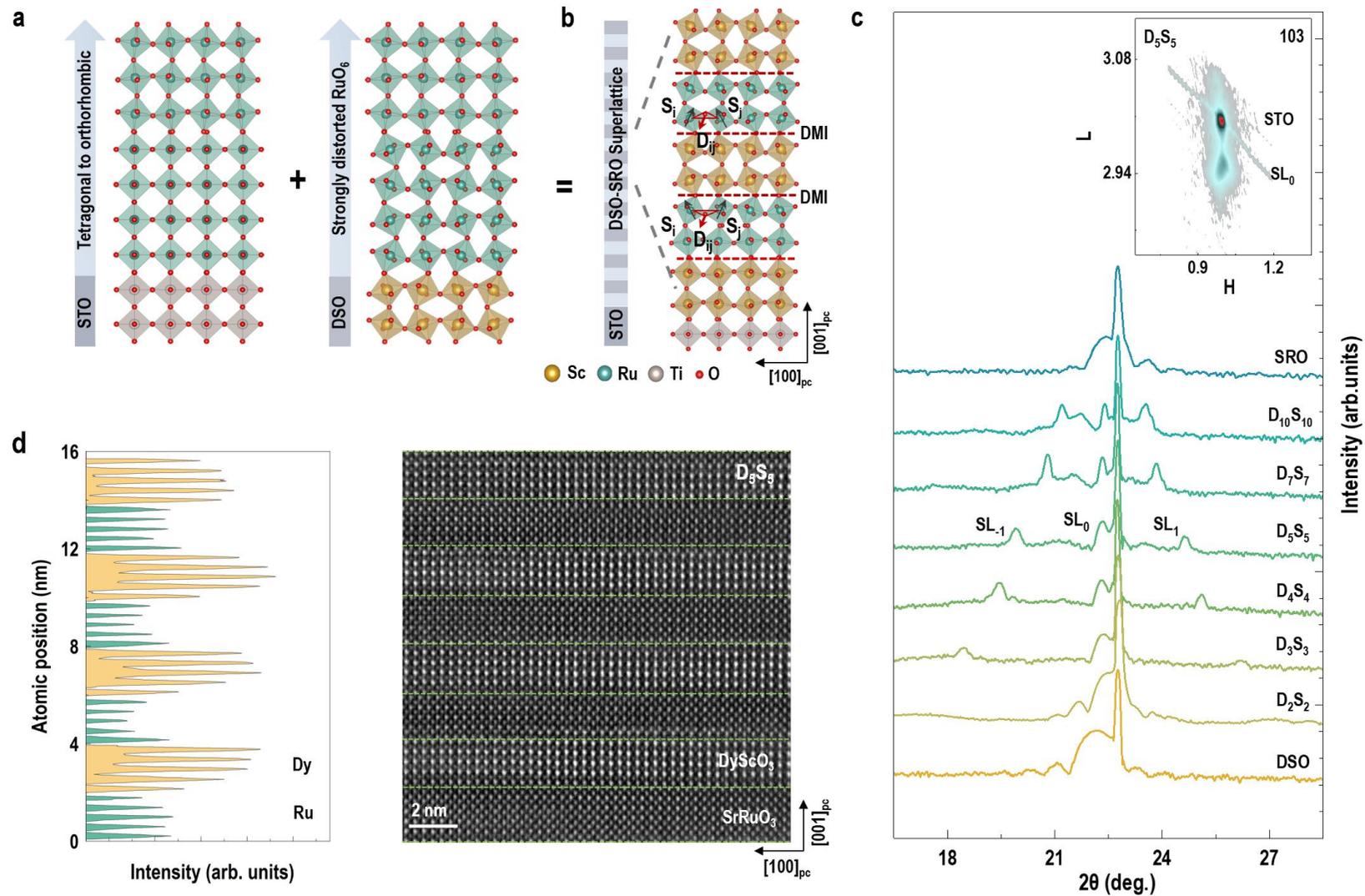

**Figure 1. Structural characterizations of the (DSO)$_n$/(SRO)$_n$ superlattices (D$_n$S$_n$), where n represents the number of unit cells of DSO and SRO layers.** (a) Schematic of RuO$_6$ octahedral structure dependent on substrate and thickness. (b) Strongly distorted RuO$_6$ octahedron induced by DyScO$_3$ crystal and epitaxial compressive strain in the artificial superlattice. (c) X-ray diffraction θ-2θ scans of D$_n$S$_n$ superlattices grown on STO substrate. The inset shows a reciprocal space map near the STO (103) reflection for the D$_5$S$_5$ sample. (d) Cross-sectional HAADF-STEM image of a representative D$_5$S$_5$ SL along the [100]$_{pc}$ zone axis. The average intensity of A-site atomic columns in the STEM image, where the yellow and green areas represent Dy and Sr atoms, respectively. The alternating DSO and SRO layers indicate the good periodicity of the superlattice.

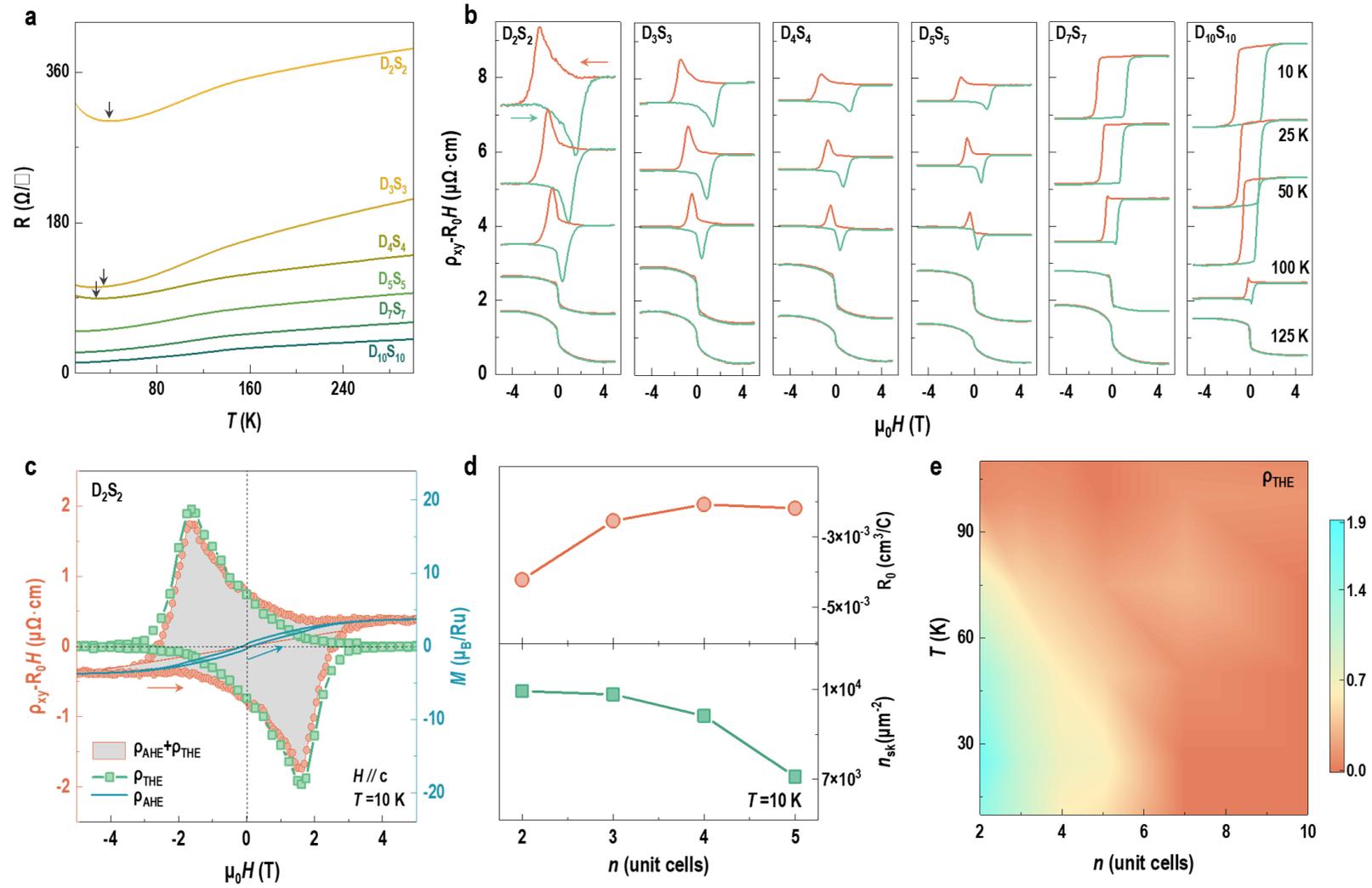

**Figure 2. Transport properties and topological Hall effects in the (DSO)$_n$/(SRO)$_n$ superlattices.** (a) Temperature-dependent resistance and (b) magnetic field-dependent Hall resistivity ($\rho_{xy}$–$R_0H$) loops obtained from D$_n$S$_n$ SLs with various SRO thicknesses ($n$) at different temperatures. The red and green lines represent opposite sweep directions of the magnetic field. The linear contribution of the ordinary Hall resistivity has been subtracted from all loops. (c) Contribution from anomalous Hall effect (AHE) and Topological Hall effect (THE) in D$_2$S$_2$ sample at $T=10$ K. The $\rho_{AHE}$ part is derived from the normalized $M$–$H$ loop. (d) Hall coefficient ($R_0$, top panel) and calculated skyrmion density ($n_{sk}$, bottom panel) for various SRO thickness at $T=10$K until topological Hall peak disappears. (e) Phase diagram of topological Hall resistivity in D$_n$S$_n$ superlattices at different temperatures.

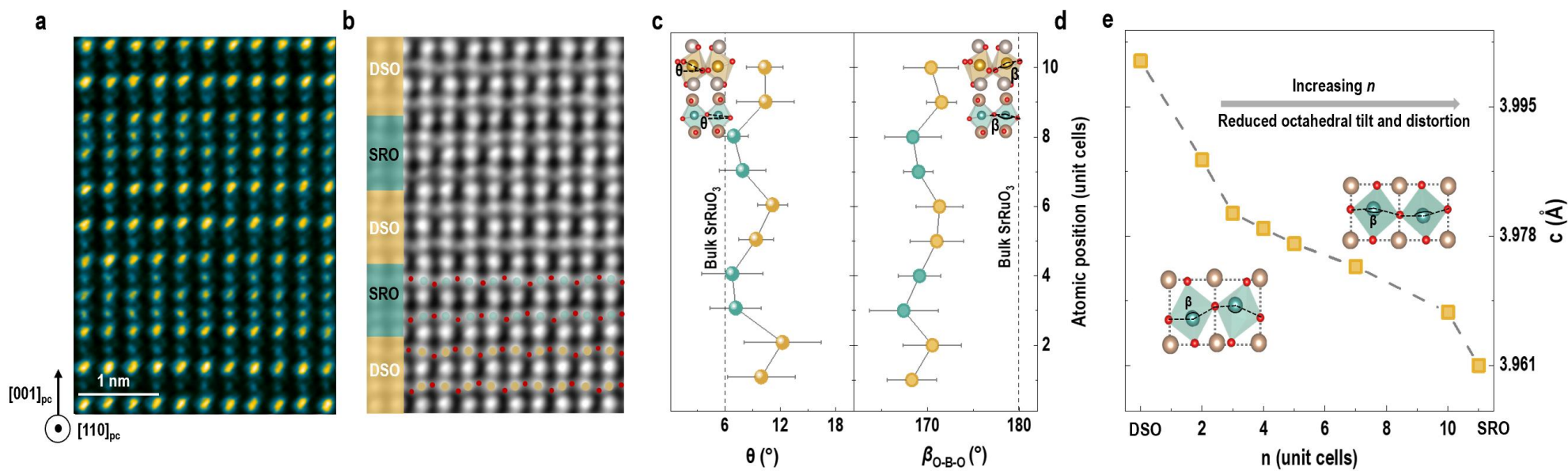

**Figure 3. Manipulation of magnetoelectric transport performance in the (DSO)$_n$/(SRO)$_n$ superlattices via octahedral distortion.** (a) Cross-sectional HAADF- and (b) ABF-STEM images of the D$_2$S$_2$ superlattice taken along the [110]$_{pc}$ zone axis. The yellow and green dots represent Dy and Ru atoms, respectively. The visualization of oxygen atoms in the ABF-STEM aids the analysis of octahedral distortion. (c) Layer position-dependent mean octahedral tilt angle ($\theta$, left panel) and (d) distorted angle between O-B-O ($\beta_{O-B-O}$, right panel), calculated from the position of all the atomic columns in (a), in which B represents Sc or Ru atom. The inserted sketches clearly denote the geometric relationships between the atoms in all octahedrons. (e) The changing tendency of c-axis lattice parameters and oxygen octahedron in the D$_n$S$_n$ superlattices with the increase of n.

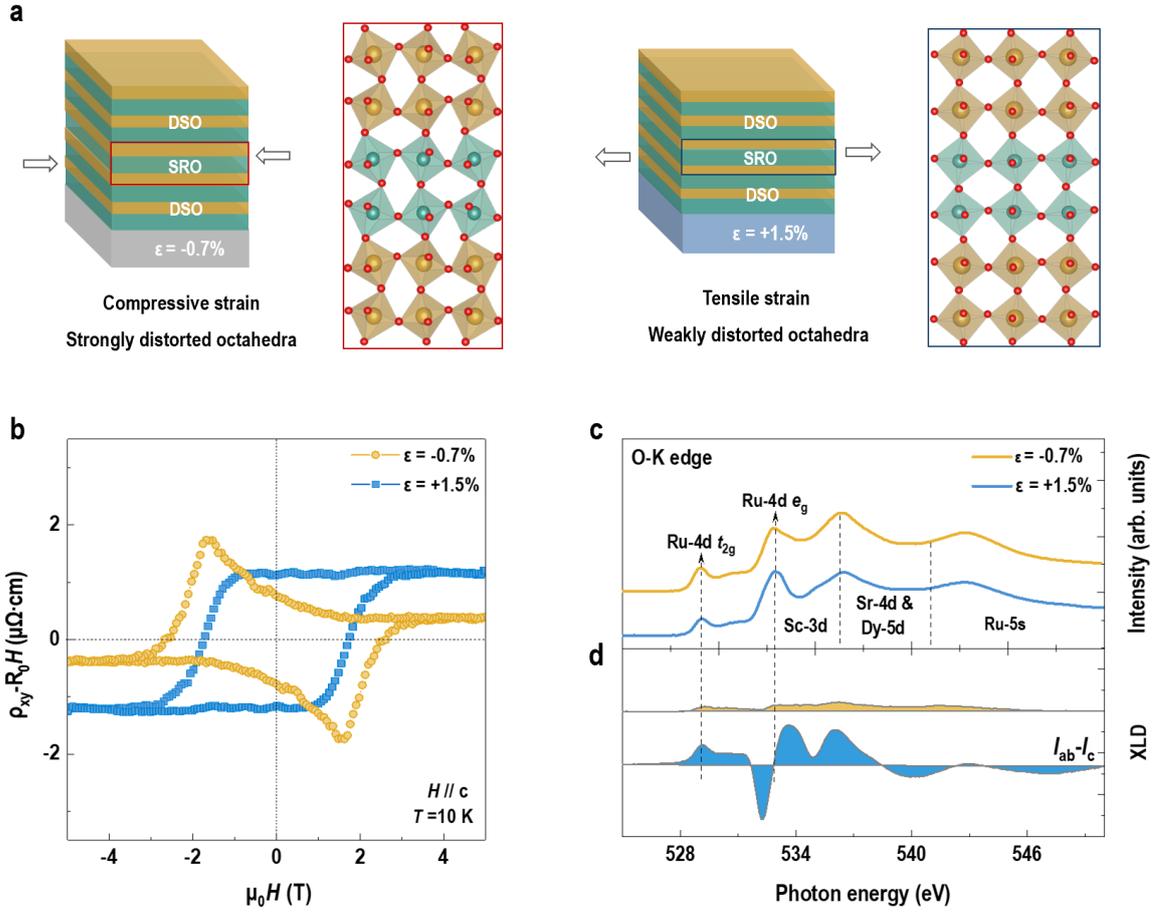

**Figure 4. Manipulation of magnetoelectric transport performance in the $(DSO)_n/(SRO)_n$ superlattices via epitaxial strain.** (a) Schematic of $RuO_6$ and $ScO_6$ octahedral distortion under compressive and tensile strain. (b) Magnetic field-dependent Hall resistivity ($\rho_{xy} - R_0H$) loops of $D_2S_2$ SLs subjected to different strain states. (c) Polarization-dependent XAS at the O K-edge for samples subjected to different strain states. (d) X-ray linear dichroism (XLD) calculated from $I_{ab}-I_c$, demonstrating the distinct orbital occupation in the SLs with different oxygen coordination.